\input{aipcheck}

\documentclass[
    ,final            
,sort&compress]{aipproc}

\layoutstyle{6x9}

\def\Msun{{M}_{\odot}}
\def\yr{\,{\rm yr}}
\def\tth{t_{\rm th}}
\def\tBV{t_{\rm BV}}
\def\tcond{t_{\rm cond}}

\newcommand{\ls}{{_<\atop^{\sim}}}

\newcommand{\gs}{{_>\atop^{\sim}}}

\begin{document}

\title{Are galactic coronae thermally unstable?}


\author{Carlo Nipoti}{
  address={Dipartimento di Astronomia, Universit\`a di Bologna, via Ranzani 1, I-40127 Bologna, Italy}
}



\begin{abstract}
  A substantial fraction of the baryons of disk galaxies like the
  Milky Way is expected to reside in coronae of gas at the virial
  temperature. This is the only realistic reservoir of gas available
  to feed star formation in the disks. The way this feeding occurs
  depends crucially on whether galactic coronae can fragment into cool
  clouds via thermal instability. Here I summarize arguments
  suggesting that galactic coronae are not prone to thermal
  instability, and I briefly discuss the implications for
  galaxy-formation models and for the origin of the high-velocity
  clouds of the Milky Way.
\end{abstract}


\maketitle


\section{Coronae of disk galaxies}

Milky Way-like galaxies are expected to be surrounded by hot rarefied
gas halos, often referred to as galactic
coronae \citep[][]{Fuk06}. These galactic coronae are hidden, in the
sense that we have only indirect evidence that they exist, so their
physical properties, such as density profile and angular momentum, are
very poorly constrained.  Indications in favour of virial-temperature
($T\sim 10^6$ K) gas around the Milky Way are, for instance, the
requirement of confinement of the high-velocity clouds (hereafter
HVCs) \citep{Spi56}, the morphology of the HVCs \citep{Bru00} and the
presence of ultraviolet absorption lines in the direction of the
HVCs \citep{Sem03}.

The Milky Way has been forming stars at a rate of $\sim 1 \Msun/\yr$
for a few gigayears, and the same rate of gas accretion is needed to
feed star formation \citep{San08}. Where does the gas feeding star
formation come from?  Accretion of cold gas from satellites is not
enough \citep{San08}, so the galactic corona is the only realistic
reservoir of gas. The question is how the hot gas is converted into
cool gas useful for star formation.  In the classical picture of
galaxy formation \citep{Whi78}, all the gas within the cooling radius
cools monolithically to form a galactic disk. However, this scenario
is not completely satisfactory, in particular because it is affected
by the so-called over-cooling problem (the predicted fraction of
condensed baryons is substantially larger than the actual fraction)
and by the angular-momentum problem (galaxies are predicted to have
too low angular momentum) \citep{Nav97}, and because it has
difficulties in reproducing the observed galaxy luminosity function
\citep{Ben03}.

In order to solve these problems, an alternative ``multi-phase
cooling'' galaxy-formation scenario has been proposed \citep{Mal04},
in which it is assumed that the hot gaseous halos of Milky Way-like
galaxies are thermally unstable, so that they do not cool
monolithically, but fragment into cool ($T\sim10^4$ K)
pressure-supported clouds.  Star formation in the disk would be fed by
these cool clouds, which, in the case of the Milky Way, should be
identified with the HVCs. Support for this scenario comes from
numerical simulations based on the smoothed particle hydrodynamics
(hereafter SPH) method \citep{Som06,Pee08}. However, given the known
difficulties of simulating with SPH a multi-phase medium with strong
density gradients \citep{Age07,Pri08}, it is debated whether the
results of these simulations are reliable or significantly affected by
numerical effects.

In fact, an independent argument, based on studies of the cooling
flows in clusters of galaxies, would suggest that galactic coronae
should not be thermally unstable. Thermal instability is the physical
process underlying the so-called ``distributed mass drop-out'' model
proposed in the attempt of solving the cooling-flow problem of
clusters of galaxies \citep{Nul86}.  The viability of the distributed
mass drop-out model was soon questioned on the basis of analytic
calculations showing that the intra-cluster medium (ICM) cannot
condense into cool clouds far from the center of the cluster, because
it is thermally stabilized by buoyancy and thermal conduction
\citep{Mal87,Bal89}.  Galactic coronae are similar in many respect to
the ICM, but have substantially lower gas temperature and density, and
might be characterized by non-negligible rotation. Provided that these
differences are taken into account, the problem of thermal instability
of galactic coronae can be studied analytically as done for the
ICM. Here I report some recent results of such analytic calculations.

\section{Thermal-stability analysis: non-rotating coronae}

\citet{Bin09} performed a thermal-stability analysis of galactic
coronae, by applying the Eulerian plane-wave perturbations approach
used by \citet{Mal87} in studying cooling flows in galaxy clusters.
The corona is modeled as a non-rotating, quasi-hydrostatic, stratified
gas in the presence of cooling and thermal conduction.  The atmosphere
is considered unmagnetized in the sense that ordered magnetic fields
are neglected (which should be a reasonable approximation, at least
far from the disk), while tangled magnetic fields have the only effect
of suppressing thermal conductivity below Spitzer's benchmark
value. An analytic dispersion relation can be obtained, so that the
linear evolution of a perturbation of given wave-vector is fully
determined by the physical properties of the corona.

Though the behavior of a given perturbation depends on the properties
of its wave-vector (tangentially-oriented perturbations have more
chances to grow than radially-oriented perturbations
\citep{Mal87,Bin09}), the stability of the system is basically
determined by three relevant timescales: the thermal instability time
$\tth$ (of the order of, but typically shorter than the cooling time),
the Brunt-V\"ais\"al\"a (buoyancy) time $\tBV$, and the
thermal-conduction time $\tcond$.  The system is unstable only when
the thermal-instability time $\tth$ is the shortest: when $\tBV$ is
shorter than $\tth$ buoyant oscillations dominate and the system is
over-stable; when $\tcond$ is shorter than $\tth$ the system is
stable.  In particular, thermal conduction damps perturbations with
wave-length $\lambda$ shorter than a critical wave-length, which is
longer for higher gas temperature.


Let us focus on the Milky Way: given the relatively poor observational
constraints available, it is possible to construct quite different
models of the Galactic corona, ranging from isothermal to adiabatic
distributions \citep{Bin09}. Applying the results of the
linear-perturbation analysis to these models \citet{Bin09} found that
an isothermal corona is not prone to thermal instability, because at
all radii $\tBV\ll\tth$, so buoyancy leads to over-stability rather
than instability; in addition short-wavelength perturbations
($\lambda\ls 10$ kpc at distances $\gs 100$ kpc from the Galactic
center) are damped by thermal conduction. Only strictly adiabatic
coronae can be unstable (the entropy profile is flat, so $\tBV\to
\infty$), but also in this case thermal conduction has a stabilizing
effect on short-wavelength perturbations. Note that damping by thermal
conduction is important even if conduction is suppressed, by tangled
magnetic fields, to $\sim\frac{1}{100}$ of Spitzer's fiducial value.

\section{Effect of rotation}

Galactic coronae might be characterized by non-negligible rotation, so
a natural question to ask is whether rotation can affect their
thermal-stability properties.  In the absence of rotation, and for
outward-increasing entropy profiles, thermal perturbations are
stabilized by buoyancy.  It is well known that there is an interplay
between convective stability and rotation, so also the evolution of
thermal perturbations might depend on rotation. In particular, if
rotation stabilizes against convection we might expect that it
destabilizes thermal perturbations, because it reduces the stabilizing
effect of buoyancy. Therefore, a first qualitative indication about
the effect of rotation on the thermal instability of galactic coronae
can be obtained by considering the results of studies of convection in
rotating stars. Almost sixty years ago, \citet{Cow51} showed that
rotation has a stabilizing effect against convection for axisymmetric
perturbations (a consequence of angular momentum conservation), but
not necessarily for non-axisymmetric perturbations: in general the
stabilizing effect of rotation is less in the latter case and it
disappears for specific perturbations. This argument suggests that
rotating galactic coronae (in the absence of thermal conduction) might
be thermally unstable against axisymmetric perturbations, but still
stabilized by buoyancy against non-axisymmetric perturbations,
provided that the distribution has outward-increasing entropy
profile. In any case thermal conduction can stabilize the
system. Preliminary results of a quantitative thermal-stability
analysis of rotating stratified atmospheres confirm these expectations
\citep{Nip10}.  For the considered problem of condensation of cool
clouds from the hot corona, we are interested in the behavior of
non-axisymmetric perturbations: in this case the results obtained in
the absence of rotation could hold also in the case of rotating
coronae.

Recently, \citet{Kau09} explored the problem of the thermal stability
of galactic coronae via high-resolution SPH-based numerical
simulation: the initial corona models have non-zero angular momentum,
but the effect of thermal conduction is not accounted for in the
simulations.  Consistent with the above analytic
results, \citet{Kau09} find that thermal instability operates only in
coronae with very shallow specific-entropy profiles, while no cool
clouds form in standard coronae with steep outward-increasing specific
entropy profiles.

\section{Conclusions}

The results reported above lead to the conclusion that multi-phase
cooling and formation of HVCs by thermal instability are unlikely to
occur in galactic coronae. In general the corona is stabilized against
thermal perturbations by buoyancy, unless it is perfectly adiabatic,
in which case it is effectively stabilized by thermal conduction.  If
thermal instability is not at work, alternative mechanisms must be
responsible for the formation of the HVCs and for feeding star
formation in the disks of galaxies like the Milky Way.  While HVCs
probably form via accretion and stripping of gas-rich satellites, the
gas responsible for feeding star formation must ultimately come from
the corona. \citet{Mar10} propose that the galactic fountain is the
link between the coronal gas and the disk: in this picture fountain
clouds grab coronal gas along their orbits via turbulent mixing, thus
continuously bringing to the disk fuel for sustaining star formation.


\begin{theacknowledgments}
I would like to thank my collaborators James Binney and Filippo
Fraternali.
\end{theacknowledgments}


\bibliographystyle{aipproc}   

\bibliography{ap001cnipoti}

\IfFileExists{\jobname.bbl}{}
 {\typeout{}
  \typeout{******************************************}
  \typeout{** Please run "bibtex \jobname" to optain}
  \typeout{** the bibliography and then re-run LaTeX}
  \typeout{** twice to fix the references!}
  \typeout{******************************************}
  \typeout{}
 }

\end{document}